\documentclass[11pt,twoside]{article}
\usepackage{jltp}
\usepackage{amsmath}
\usepackage{amssymb}
\usepackage{bm}

\newcommand{\kk}{{\bf k}}
\newcommand{\rr}{{\bf r}}

\usepackage{graphicx}

\title{Universality in Voltage-driven Nonequilibrium Phase 
Transitions\footnote{Dedicated to Prof. Hilbert von L\"ohneysen on the
occasion of his 60th birthday.}}

\author{Johann Kroha\address{Physikalisches Institut, 
Universit\"at Bonn, Nussallee 12, 53115 Bonn, Germany},
Michael Arnold, and Beate Griepernau\footnote{Present address: Center for
Bioinformatics, Universit\"at des Saarlandes, P.O. Box 151150, 
66041 Saarbr\"ucken, Germany}}

\runninghead{J. Kroha, M. Arnold, and B. Griepernau}
            {Universality in Nonequilibrium Phase Transitions}
       
\begin{document}

\begin{abstract}
We consider the non-equilibrium 
ferromagnetic transition of a mesoscopic sample of a resistive
Stoner ferromagnet coupled to two paramagnetic leads. The
transition is controlled
by either the lead temperature $T$ or the transport voltage 
$V$ applied between the leads. We calculate the $T$ and $V$ 
dependence of the magnetization. For systems
with a flat density of states we find within mean-field theory
that even at finite bias the
magnetization does not depend on the position along the sample
axis, although the charge density and other quantities do vary.
This may be relevant for possible spintronics applications.
In addition, we establish a generalized control parameter in terms
of $T$ and $V$ which allows for a universal description of the
temperature- and voltage-driven transition.  

\end{abstract}

\maketitle


\section{INTRODUCTION}
Phase transitions in interacting electron systems far from thermodynamic 
equilibrium are an exciting, relatively young field of research, both for 
fundamental theoretical reasons and for application purposes.
While the theory of equilibrium phase transitions is well developed, a
theory for phase transitions out of equilibrium is much more difficult to 
formulate, largely because in the most general case the thermodynamic states 
of the system are not characterized by a minimum of the free energy and 
because often the temperature $T$ as a state variable driving the 
transition is not even defined. On the other hand, nonequilibrium 
phase transitions are of high technological potential, since a mesoscopic 
electron system can be driven out of equilibrium in a highly controlled way,
and controlling a nonequilibrium phase transition by externally applied 
fields can be envisaged. 
It opens up the prospect of fast switching between different 
magnetic or conducting states of a mesoscopic system, e.g. by applying
a transport bias voltage or by an applied laser field, 
with possible spintronics applications.

A thermodynamic phase transition {\it in equilibrium} 
is in general controlled by the minimum of the free energy 
\begin{equation}
{\cal F}={\cal U}-  T S \ ,
\label{eq:F}
\end{equation}
i.e. by the balance of the internal energy ${\cal U}$ and the 
entropy $S$. The dependence on the control parameter temperature
favors an ordered phase at low $T$ and a disordered phase at
high $T$. In a {\it stationary nonequilibrium situation} 
internal energy and entropy are statistically still well-defined
quantities. One may, therefore,  
conjecture that a relation analogous to Eq.\  \ref{eq:F} holds even
in this case. The different phases of a stationary non-equilibrium 
system would then again be determined by the minimum of the free energy.
However, in the above relation (\ref{eq:F}) the entropy coefficient $T$ 
would have to be generalized, since temperature $T$ is in general not
defined in nonequilibrium. Based on general thermodynamic arguments we 
propose in the present article a definition of such a 
generalized control parameter $\tau$ which allows to describe 
electron systems driven out of equilibrium
by an applied stationary bias voltage $V$ in a unified way.   

\begin{figure}[t]
\begin{center}
\includegraphics[width=0.9\linewidth]{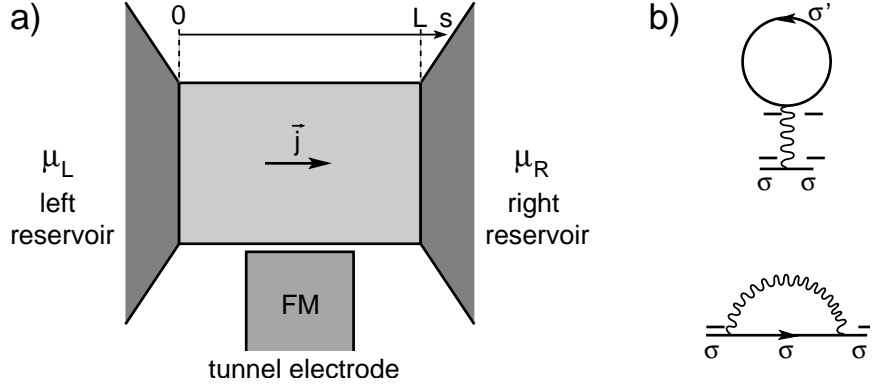}
\end{center}
\vspace*{-0.4cm}
\caption{\label{fig1} 
a) Schematic view of a magnetic transitor. The ferromagnetic 
mesoscopic sample is shown in light grey.
It is attached to paramagnetic leads kept at chemical potentials
$\mu_L$ and $\mu_R$, respectively. A magneto-tunneling current 
depending on the sample magnetization can be extracted through
the third electrode FM. The position coordinate 
along the sample of length $L$ is denoted by $s$, as shown in the figure. 
b) The Keldysh Hartee-Fock diagrams defining the Stoner 
mean-field theory in stationary nonequilibrium. Solid lines denote
the electron propagators, wavy lines the screened Coulomb interaction
$U$.
}
\end{figure}

To be specific, we consider the voltage-driven 
magnetic phase transition of a Stoner ferromagnet. Such a system 
can be realized as a mesoscopic sample of a magnetic metal, contacted 
by two paramagnetic electron reservoirs or leads at chemical potentials 
$\mu_L$ and $\mu_R$, respectively, as seen in Fig.\ \ref{fig1}a). 
It has recently been demonstrated that the fabrication of such 
devices, which sustain a high nonequilibrium bias, is technologically 
feasible using metallic nanobridges with high lead-to-bridge aspect ratio 
\cite{weber01} or using resistive nanowires.\cite{pothier97}
As will be seen below, 
the ferromagnetic transition in the mesoscopic sample can be controlled 
both by the reservoir temperature $T$ and by the voltage $V=\mu_L-\mu_R$ 
applied between the reservoirs. The magnetic state of the sample
may be probed by a third, ferromagnetic electrode, attached to the sample 
via a tunnel junction. This electrode stays in equilibrium with itself, 
independent of the transport voltage $V$, so that it remains in the 
ferromagnetically ordered state throughout the transition of the sample,
and its tunneling conductance depends strongly on the magnetization
in the sample.
A related proposal of controlling magnetism by a nonequilibrium 
transport bias has recently been put forward in Ref.\ [3].
However, in that proposal the magnetization was induced in a 
paramagnetic metal sample by ferromagnetism in the leads, and the        
sample was assumed to be zero-dimensional in the sense that it was 
smaller than the electrons' elastic mean free path 
$\ell$, implying ballistic transport and a position independent electron
distribution function in the metal sample.
In the present work we consider the more realistic case of a 
resistive, ferromagnetic sample whose lateral size $L$ 
is greater than the elastic mean free path, $L>\ell$, 
but still smaller than the inelastic
and spin relaxation lengths, $L<\ell_{in}, \ell_s$. In this case the 
transport is diffusive, leading to a nonequilibrium electron
distribution function which depends on the location along the sample. 
We develop a mean field theory for the temperature- or voltage-driven Stoner 
transition. 

One might think that this transition, driven by the transport voltage 
at reservoir temperature $T=0$ might be a quantum phase transition,
since it is controlled by a parameter different from the temperature. 
However, at any finite voltage $V\neq 0$ the nonequilibrium quasiparticle
distribution is such that there exists a finite phase space for inelastic
scattering (see, e.g., Section 2.1), so that the quantum dephasing rate 
in the sample, $\gamma_{\varphi}$,  
remains finite even for reservoir temperature $T=0$. Thus, 
critical divergencies of correlation functions at the phase transition
are dominated by classical, statistical fluctuations rather than quantum
fluctuations, i.e. the transition belongs to the universality class of
classical rather than quantum phase transitions, in contrast to the 
assertion of Ref.~[4]. Remarkably, the authors of Ref.~[4] do, however, 
find that the transition is governed by the classical critical exponents. 
In fact, as will be seen in Section 3, the roles of $V$ and $T$ are 
completely analogous, as far as thermodynamic properties are concerned.  
 
This paper is organized as follows. 
In Section 2 we define the Stoner model in a steady-state
nonequilibrium situation with finite bias. 
The quantum Boltzmann equation for a resistive wire
with a Stoner mean-field interaction is derived in Section 2.1, 
and the Stoner mean-field equations in the presence of finite bias 
are presented in Section
2.2, along with their numerical solution for various temperatures 
and voltages. In Section 3 we provide general thermodynamic arguments
which allow us to introduce a control parameter $\tau (T,V)$,
depending on the reservoir temperature $T$ and the bias voltage $V$,
as a generalization of the temperature to the voltage-biased
non-equilibrium situation, as far as thermodynamic properties are
concerned. We demonstrate explicitly that this provides a 
universal description of the voltage-driven 
Stoner transition. Some concluding remarks are given in Section 4.

\section{VOLTAGE-DRIVEN STONER FERROMAGNETISM}

As a simple model for a ferromagnetic transition in an
itinerant, diffusive electron system we consider electrons
on a lattice with dispersion $\varepsilon _{\kk}$, with a
random potential $V_j$ induced by static impurities on lattice
sites $j$ and a strongly screened, local Coulomb repulsion 
$U$ between electrons on the same site,
\begin{equation}
H=\sum_{{\kk}\sigma}\varepsilon_\kk 
         c^{\dagger}_{\kk\sigma}c^{\phantom{\dagger}}_{\kk\sigma}
+\sum_{j\sigma} V_j c^{\dagger}_{j\sigma}c^{\phantom{\dagger}}_{j\sigma}
+U\sum_j c^{\dagger}_{j\uparrow}c^{\phantom{\dagger}}_{j\uparrow}
         c^{\dagger}_{j\downarrow}c^{\phantom{\dagger}}_{j\downarrow}\ .
\label{eq:hamiltonian}
\end{equation}
Here $c^{\dagger}_{\kk\sigma}$, $c^{\phantom{\dagger}}_{\kk\sigma}$
denote the electron creation and destruction operators with 
momentum $\kk$ and spin $\sigma=\pm 1/2$, and 
$c^{\dagger}_{j\sigma} = \sum_{\kk} {\rm exp}(i\kk {\bf x}_j) 
c^{\dagger}_{\kk\sigma}$ is the creation operator at site $j$. 
We assume that any antiferromagnetic instabilities that could be 
induced by nearest neighbor hopping on a cubic lattice 
are frustrated, e.g., by an appropriate lattice structure or by a
next-nearest neighbor hopping included in the band structure 
$\varepsilon _{\kk}$.  

\subsection{Nonequilibrium Distribution in a Diffusive Ferromagnet}

Any theory of an interacting system out of equilibrium requires 
knowledge of the quasiparticle distribution $f_{\sigma x}(\omega)$
as a function of the particle energy $\omega$. 
In a paramagnetic, diffusive system at bias $V$, if inelastic relaxations can 
be neglected, $f_{\sigma x}(\omega)$ is known 
\cite{nagaev92,nagaev95,kozub95,pothier97} 
to have a double-step form according to the two Fermi edges in 
the reservoirs, and to depend linearly on the position $x=s/L$ 
along the sample axis (see Fig.\ \ref{fig1}a)). 
In a ferromagnet, however, the Stoner splitting may in general 
depend on $x$ as well, inducing nontrivial position and spin
dependences to $f_{\sigma x}(\omega)$
even in the absence of inelastic or dynamical spin flip processes.
In the following we derive the distribution function for the case
of a mesoscopic ferromagnet, where ineleastic and spin flip processes
can be neglected.  
 
The kinetic equation (quantum Boltzmann equation) for $f_{\sigma x}(\omega)$ 
is obtained as the off-diagonal part of the Keldysh Dyson equation
for the lesser Green's function $G_{\sigma}^{<}(\omega, 
{\bf x}, {\bf x}')$.\cite{landau} 
After transforming to center-off-mass and relative coordinates, 
${\rr}=({\bf x}+{\bf x}')/2$, $\Delta {\bf x}={\bf x}-{\bf x}'$,
taking $\Delta {\bf x}$ to be confined in a small volume 
compared to the scale on which the external fields vary, 
and Fourier transforming to momentum space with respect to $\Delta {\bf x}$
within that volume, the resulting quantum Boltzmann equation acquires
the form of a continuity equation in phase space,
\begin{equation}
\left[
\frac{\partial}{\partial t} +   {\bf v}_{\kk} \cdot \nabla _r
 + \frac{-e}{\hbar} {\bf E}(\rr ) \cdot \nabla _k \right] 
G^<_{\sigma{\rr}}(\kk , \omega) ={\cal I}\left\{ f_{\sigma \rr}(\kk) \right\}
\label{eq:boltzmann}
\end{equation}
where $G^<_{\sigma\rr}(\kk , \omega) = 2\pi i f_{\sigma \rr}(\kk) 
A_{\sigma\rr}(\kk,\omega) $ is expressed in terms of the 
distribution $f_{\sigma \rr}(\kk)$ and the spectral density 
$A_{\sigma\rr}(\kk,\omega)$ of momentum $\kk$ at position $\rr$ in the
sample. ${\bf v}_\kk  = {\partial \rr}/{\partial t} = 
\partial \varepsilon _\kk / \partial \kk$ is the electron 
group velocity and $-e {\bf E(\rr)}=\hbar {\partial \kk}/{\partial t}$
the force acting on the electrons due to an external field ${\bf E}(\rr)$.
The latter includes both, the external bias voltage and the 
electrostatic potential generated by the random impurities.
Since we consider a stationary situation and assume that the sample
is smaller than the inelastic and the spin relaxation lengths, the explicit 
time derivative and the collision integral 
${\cal I}\left\{ f_{\sigma \rr}(\kk) \right\}$ 
vanish in Eq.\ (\ref{eq:boltzmann}). 
After summation over 
$\kk$ the force term vanishes as well. This is because of the 
relation 
\begin{equation}
\sum _{\kk\sigma} \nabla _k\cdot {\bf E}(\rr ) f_{\rr}(\kk) 
A_{\sigma\rr}(\kk,\omega)=
\oint _{S_k} {\bf E}(\rr ) f_{\sigma\rr}(\kk) A_{\sigma\rr}(\kk,\omega) 
\cdot d{\bf S}_{\kk} = 0 \ .
\label{eq:ksum}
\end{equation}
Here
the second term implies an integration over the surface of the 1st 
Brillouin zone, where $f_{\sigma\rr}(\kk)$ is either 0 or 1, i.e.
$f_{\sigma\rr}(-\kk)=f_{\sigma\rr}(\kk)$ and 
$d{\bf S}_{-\kk}=-d{\bf S}_{\kk}$. The same $\kk$-summation 
also introduces the particle and the current density
in Eq.\ (\ref{eq:boltzmann}),
\begin{eqnarray}
\rho_{\sigma\rr}(\omega) &=&
\sum_{\kk}f_{\sigma\rr}(\kk)A_{\sigma\rr}(\kk,\omega)
\label{eq:density}\\
{\bf j}_{\sigma\rr}(\omega) &=&
\sum _{\kk}{\bf v}_\kk f_{\sigma\rr}(\kk)A_{\sigma\rr}(\kk,\omega) \ ,
\label{eq:current}
\end{eqnarray}
which in a diffusive system with diffusion coefficient $D$ 
are related by Fick's law,
\begin{eqnarray}
{\bf j}_{\sigma\rr}(\omega) &=& - D \nabla _r \rho_\rr (\omega) \ .
\label{eq:fick}
\end{eqnarray}
In this way the diffusive Boltzmann equation in a ferromagnet takes
the form
\begin{equation}
- D \nabla_r^2 \left[f_{\sigma\rr} (\omega)N_{\sigma\rr}(\omega)\right] = 0 \ ,
\label{eq:diff_boltzmann}
\end{equation}
where $N_{\sigma\rr}(\omega) = \sum_{\kk}A_{\sigma\rr}(\kk,\omega)$ is
the Stoner split electronic density of states 
(DOS) at position $\rr$ in the sample.  
Note that in a paramagnetic system $N_{\sigma\rr}(\omega)$
is in general position independent and can be dropped from Eq.\
(\ref{eq:diff_boltzmann}). In a ferromagnet, however, the spectral
function contains the Stoner selfenergy $\Sigma _{\sigma\rr}$ 
(see Section 2.2), which depends in general on the 
position $\rr \equiv s$,
\begin{equation}
A_{\sigma x}(\kk,\omega) = -\frac{1}{\pi}{\rm Im} \frac{1}
{\omega+\mu_0-\varepsilon_{\kk}-\Sigma _{\sigma x}+i0} \ ,
\end{equation}
where $s$ is the coordinate along the sample axis (see Fig.\ref{fig1}a))
and $x=s/L$. Taking into account only the linear dependence of the DOS on
$x$, with
$N'_{\sigma}(\omega)\equiv dN_{\sigma x}(\omega)/dx|_{x=0.5}=
-[dN_{\sigma x}(\omega)/d\omega\cdot d\Sigma_{\sigma x} /dx ]|_{x=0.5}$ and 
$N_{\sigma}(\omega)\equiv N_{\sigma x}(\omega)|_{x=0.5}$,  
the quantum Boltzmann equation in a resistive ferromagnet takes the
final form
\begin{equation}
- D \frac{d^2 f_{\sigma x}(\omega)}{dx^2} 
- 2 D \ \frac{N'_{\sigma}(\omega)}{N_{\sigma}(\omega)}\ 
  \frac{d f_{\sigma x}(\omega)}{dx} =0 \ .
\label{eq:fm_boltzmann}
\end{equation}
The second term in this equation implies a selfconsistent coupling 
to the Stoner mean-field equations 
via the spin and position dependent DOS. 
However, Eq.\ (\ref{eq:fm_boltzmann}) can readily be solved analytically, 
once $N'_{\sigma}(\omega)$ and $N_{\sigma}(\omega)$ are determined.
This will be done in a forthcoming publication.\cite{arnold06}
In the present paper we wish to focus on universality related issues,
which can be discussed as well for the particle-hole symmetric case.
Therefore, we will assume in the following that the DOS is flat 
within the energy window defined by the applied bias voltage $V$.
In this case $N'_{\sigma}(\omega)=0$, and
the kinetic equation (\ref{eq:fm_boltzmann}) reduces to
the form familiar from noninteracting diffusive systems. In particular,
$f_{\sigma x}(\omega)$ is then spin independent, since the boundary
conditions on $f_{\sigma x}(\omega)$ are spin symmetric, i.e. both 
spin species in the sample couple to the same paramagnetic reservoirs.
\begin{equation}
f_{\sigma x} (\omega) = f_{-\sigma x} (\omega) \equiv f_{x} (\omega) 
\qquad {\rm for}\  N'_{\sigma}(\omega)=0 \ .
\end{equation}
The Boltzmann equation (\ref{eq:fm_boltzmann}) is easily solved 
for this case with the boundary conditions that in the left (right)
lead there is an equilibrium distribution with chemical potenial
$\mu_L=+eV/2$ and $\mu_R=-eV/2$, respectively,
\begin{equation}
f_x(\omega) = x f^{(0)}(\omega+\frac{eV}{2}) + 
          (1-x) f^{(0)}(\omega-\frac{eV}{2}) \ ,
\label{eq:fx}
\end{equation}
where $f^{(0)}(\omega)= 1/[{\rm exp}(\omega/k_BT)+1]$ is the 
Fermi distribution and $T$ the reservoir temperature. 

\subsection{Stoner Mean-Field Equations in Stationary Nonequilibrium}

For simplicity we assume an
elliptic DOS (per spin) of the disordered system without interactions, 
$N_0(\omega +\mu_0 ) = (2/\pi)\sqrt{1-(\omega +\mu_0)^2}$, where
the single-particle energy $\omega$ is measured 
relative to the equilibrium chemical potential $\mu_0$, and all
energies are given in units of the half bandwidth $B$.
To realize a flat DOS for energies within the voltage window, 
$-eV/2\leq \omega \leq eV/2$, $\mu_0$ is placed at the band
center, $\mu_0=0$, and $eV\ll B$.
As discussed in Section 2.1, the distribution function is then 
decoupled from  the solution of the
Stoner model and is given by Eq.\ (\ref{eq:fx}). 

The Hartree-Fock selfenergy for this model is represented by
the Keldysh diagrams shown in Fig.\ \ref{fig1}b). 
Note that all vertex points of the instantaneous interaction $U$
are on the same branch of the Keldsyh contour, as indicated by the
``$-$'' signs, and that the equal-time operator products are
normal-ordered. The interal equal-time propagators are, therefore,
the lesser Green's functions $G^<(\omega)$.\cite{landau} 
The Hartree-Fock selfenergy then reads,
\begin{eqnarray}
\Sigma _{\sigma} (T,V,x) &=&
U\langle n_{-\sigma} \rangle\nonumber \\ &=& 
U \int d\omega
f_x(\omega) N_0[\omega +\mu_0 - \Sigma_{-\sigma}(T,V,x)] \ ,
\label{eq:HF1}
\end{eqnarray}
where $\langle n_{\sigma} \rangle$ is the conduction electron number
per site with spin $\sigma$.
At $T=0$ the integral can be evaluated, and the $V$ and $x$ dependence can be
made explicit, 
\begin{eqnarray}
\Sigma _{\sigma} (0,V,x) &=& 
\frac{U}{\pi} \left[
  x \left( \Omega_{-,-\sigma} \sqrt{1-(\Omega_{-,-\sigma})^2}
        +{\rm arcsin}(\Omega_{-,-\sigma})
  \right)\right. \label{eq:HF2}\\ &+&
\left.(1-x) \left( \Omega_{+,-\sigma} \sqrt{1-(\Omega_{+,-\sigma})^2}
        +{\rm arcsin}(\Omega_{+,-\sigma})
  \right) + \pi
\right] \, , \nonumber
\end{eqnarray}
with $\Omega_{\pm,\sigma}=
\pm\frac{eV}{2} +\mu_0 - \Sigma_{\sigma}(0,V,x)$.
The selfenergy can be separated in a spin independent and a 
spin dependent part, $\Sigma _{\sigma} = \Sigma _{0}+ \sigma \Delta \Sigma$,
with
\begin{eqnarray}
\Delta \Sigma  &=&\Sigma _{\uparrow} - \Sigma _{\downarrow} 
\label{eq:DSigma}\\
\Sigma _0 &=& \frac{1}{2}
\left(\Sigma _{\uparrow} + \Sigma _{\downarrow} \right) = \frac{U}{2}
\langle n \rangle \ _x .
\label{eq:Sigma0}
\end{eqnarray}
\begin{figure}[t]
\begin{center}
\includegraphics[width=0.9\linewidth]{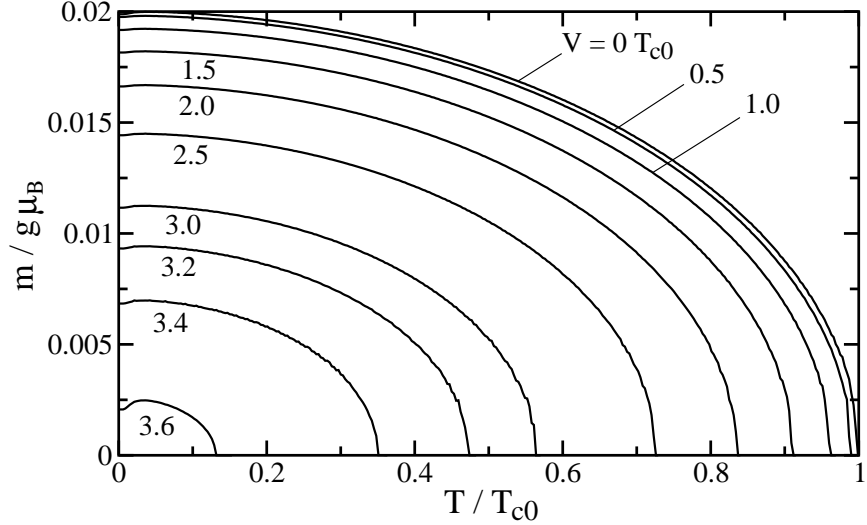}
\end{center}
\vspace*{-0.4cm}
\caption{\label{fig2} 
The magnetization per site is shown as a function of $T$ 
for various values of the transport voltage $V$ as indicated. 
$V$ is given in units of the equilibrium Curie temperature
$T_{c0}$. The screened Coulomb interaction was chosen such
that $T_{c0}=0.01\ B$ (i.e. $UN_0(0)=1.0001669$). 
The small dips in the $m(T)$ curves near $T=0$  
are a consequence of the slightly non-constant
DOS for high bias $V$.
}
\end{figure}
It is now crucial that the definition (\ref{eq:fx}) of the
nonequilibrium distribution fixes the zero of the energy to be in the
middle between the two Fermi steps {\it for each position x along the
sample}. Therefore, in Eqs.\ (\ref{eq:HF1}), (\ref{eq:HF2}) the 
interaction-induced, spin independent shift of the energy 
$\Sigma_0$ must be absorbed into the (equilibrium)
chemical potential {\it for each x}: 
$\mu_0 -\Sigma _0 \longrightarrow \mu_0$.
The selfenergy then consists only of the spin dependent part, 
$\Sigma_{\sigma}\equiv \sigma\Delta \Sigma$. Eq.\ (\ref{eq:DSigma}), 
together with Eqs.\ (\ref{eq:fx}) and
(\ref{eq:HF1}) constitute the unconstrained, selfconstistent 
Hartree-Fock equations
which determine the magnetization per site, $m=g\mu _B \Delta \Sigma /U$,
$g$ and $\mu_B$ being the Land\'e factor and the Bohr magneton,
respectively.
We note that for the present case of a flat DOS it follows immediately
from Eqs.\ (\ref{eq:HF1}), (\ref{eq:DSigma}), (\ref{eq:Sigma0})  
and from the linearity of $f_x(\omega)$ with respect to $x$, 
Eq.\ (\ref{eq:fx}),
that (i) $\Delta\Sigma$, and hence
the magnetization, are independent of the position $x$ and 
that (ii) the total particle 
density $\langle n \rangle_x$ is a linear function of $x$. The latter
is consistent with Fick's law (\ref{eq:fick}) and the fact that 
the stationary current ${\bf j}_x$ in the sample is homegenous.
The $x$ independence of the magnetization $m$ remarkably implies that
the total sample magnetization can be switched {\it as a whole} 
by a transport voltage $V$, although at finite bias the electron density 
$\langle n \rangle _x$ and other quantities vary along the sample.
This may be of technological relevance for spintronics devices, although
this property does not persist exactly for non-constant DOS.\cite{arnold06}
The mean-field solutions for the magnetization $m$ as a function of 
reservior temperature $T$ are shown for various applied transport voltages $V$
in Fig.\ \ref{fig2} and as a function of $V$ for various $T$ in 
Fig.\ \ref{fig3}. For these data the strength of the screened 
Coulomb interaction $U$ has been chosen such that the Stoner criterion,
$UN_0(0)>1$, is satisfied and that the equilibrium  Curie temperature is 
$T_{c0}=0.01\ B$, corresponding to $T_{c0}\approx 300\ K$ for a 
half bandwidth of $B=3\ eV$.

\begin{figure}[t]
\begin{center}
\includegraphics[width=0.9\linewidth]{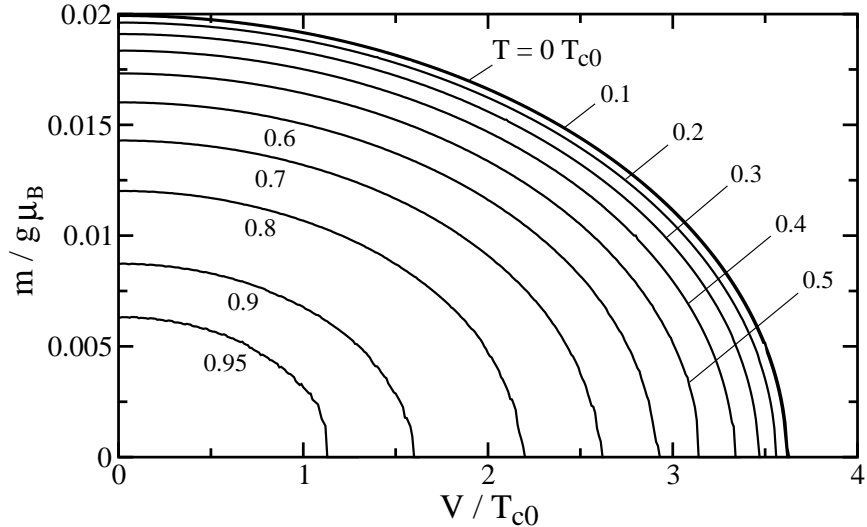}
\end{center}
\vspace*{-0.4cm}
\caption{\label{fig3} 
The magnetization per site is shown as a function of $V$ 
for various values of $T$, as indicated. 
The parameter values used are as in Fig.\ \ref{fig2}. 
}
\end{figure}

\section{UNIVERSALITY}

The fact that the Stoner transition can be controlled by both,
temperature $T$ and transport voltage $V$, the similarity of 
the magnetization curves $m(T)$ and $m(V)$ in Figs.\ \ref{fig2}
and \ref{fig3}, and the independence of the magnetization of
the position $x$ in the sample for a flat DOS suggest that 
there is a universality in this transition in terms of the
two control parameters $T$ and $V$. Indeed, the complete
thermodynamics of any system is determined, once the internal
energy ${\cal U}$ or the free energy ${\cal F}$
(with appropriate state variables) is known. At least the 
internal energy is well-defined also in a stationary 
nonequilibrium situation. For any {\it stationary} electron system 
in or out of equilibrium, ${\cal U}$ 
is uniquely determined by the quasiparticle 
distribution function $f_x(\omega)$. The latter is, in turn,
uniquely defined by the applied transport voltage $V$ (and the 
concomitant boundary conditions), if the system is coupled to 
one or more reservoirs to ensure stationarity, 
and by the reservoir temperature $T$. This suggests that it should be
possible to define a single, unified state variable, involving 
both $T$ and $V$, in terms of the internal energy.    

Therefore, we put forward the following procedure. 
The temperature of an equilibrium system is usually measured by
bringing it in heat contact with a reference system whose internal 
energy is known as a function of $T$, like an ideal gas 
(ideal gas thermometer). Since the ideal gas is not suitable for
applying a transport voltage, we propose to use a diffusive,
noninteracting electron gas as a reference system for any 
voltage-biased, interacting system, like, e.g., a mesoscopic
Stoner ferromagnet, Eq.\ (\ref{eq:hamiltonian}). 
As before, we will restrict ourselves here to the case that 
the DOS is constant within the voltage window, $N_0(0)$, and 
defer the general case to a forthcoming paper.\cite{arnold06} 
The change of the 
internal energy density of this reference system as a function of 
$T$ and $V$ is well-defined and is readily calculated as,
\begin{eqnarray}
\Delta {\cal U} &=& {\cal U}(T,V)- {\cal U}(0,0)\nonumber \\ 
&=& N_0(0)
\int _{-\infty}^{\infty} d\omega\ \omega 
\left[ f_x (\omega ,T,V) - f^{(0)} (\omega ,0,0)  
\right] \label{eq:DeltaU}\\
&=&\frac{\pi^2}{6}\, N_0(0)\,\left[ 
(k_BT)^2 + \frac{3}{4\pi^2} (eV)^2
\right] \nonumber \ .
\end{eqnarray}
\begin{figure}[t]
\begin{center}
\includegraphics[width=0.9\linewidth]{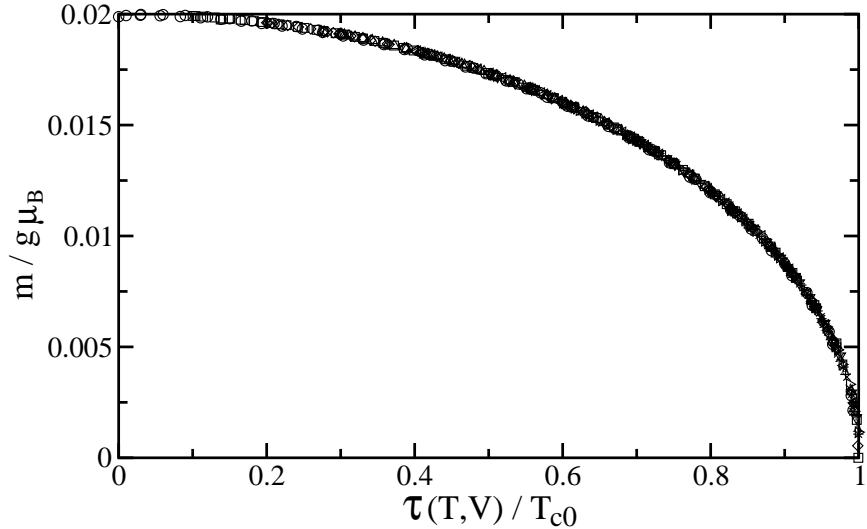}
\end{center}
\vspace*{-0.4cm}
\caption{\label{fig4}
Collapse of all magnetization data contained in 
Figs.\ \ref{fig2} and \ref{fig3} onto a single curve, depending
on $\tau (T,V)$ (Eq.\ \ref{eq:tau}) only. The data for 
different parameter values of $T$ and $V$ are displayed here
using different symbols.
}
\end{figure}
Remarkably, $\Delta {\cal U}$ does not depend on the position $x$
where the energy density is considered, as long as the DOS is 
flat, in accordance with the results of Section 2.2. 
The expression (\ref{eq:DeltaU}) can now be used to 
extract a generalized state variable, $\tau(T,V)$, that uniquely determines 
the internal energy of the reference system, and thus the 
thermodynamics of any stationary electron system controlled by the state 
variables $T$ and $V$,
\begin{eqnarray}
\tau(T,V) = \sqrt{T^2 + \frac{1}{4{\cal L}}V^2} \ ,
\label{eq:tau}
\end{eqnarray}
where ${\cal L}=\pi^2k_B^2/3e^2$ is the Lorentz number, 
and $\Delta {\cal U} = ({\pi^2}/{6}) N_0(0) \tau^2$. 
Note that the form of the expression for $\tau(T,V)$ is equivalent
to the effective temperature of a voltage-biased, hot electron system that has
equilibrated due to strong inelastic processes.\cite{nagaev95}
However, in our line of arguments above no assumption has been made about 
whether or not there are inelastic processes involved. This supports the
universality of the definition Eq.\ (\ref{eq:tau}). 

We have tested whether the Stoner transition can be described by 
$\tau(T,V)$ in a universal way. In Fig.\ \ref{fig4} we
have plotted all the data contained in Figs.\ \ref{fig2} and \ref{fig3} 
as a function of $\tau(T,V)$. Indeed there is a perfect collapse of all
data onto a single curve, as expected.

\section{CONCLUSION}

Even the simple Stoner model of a ferromagnetic transition
can exhibit interesting behavior when considered in a non-equilibrium 
situation. We have used this model to define a generalized
control parameter $\tau (T,V)$ which allows to describe the
voltage and temperature dependence of the Stoner transition 
in a unified way. The model has been evaluated within mean-field theory
where the critical exponent of the magnetization in dependence of 
both $T$ and $V$ is $\alpha = 1/2$. 
We emphasize that the generality of our thermodynamic arguments 
that led to the definition of $\tau(T.V)$ 
strongly suggest that it not only allows for a 
universal description of the magnetization but also of the complete
thermodynamics of the Stoner model, and that this holds true even
beyond mean field theory. In particular, correlations will certainly
change the value of $\alpha$, but it will remain the same for 
the $T$ and the $V$ dependence. Moreover, the thermodynamic 
arguments suggest that not only Stoner ferromagnets but 
the thermodynamics of any interacting electron system controlled by 
the reservoir temperature $T$ and a nonequilibrium transport voltage $V$ 
can be described by $\tau (T,V)$ in a universal way.
Conversely, it is obvious that dynamical and spectral 
properties are influenced by $T$ and $V$ in quite different ways,
which cannot be captured only by the $T$ and $V$ dependence of the
internal energy. Therefore, no unified behavior is expected for
these quantities. 

As mentioned above, the magnetization $m$ will acquire an $x$-dependence, as
soon as the density of states in the sample as function of energy 
is not flat. In this case the interaction-induced correlation length
will also influence the spatial dependence of $m$.       
The thermodynamics of voltage-driven systems, the influence of 
correlations in general and the position dependence of the
magnetization in particular will be subjects of further research.

\section*{ACKNOWLEDGMENTS}
This work is supported by DFG through SFB 608 and through
KR1726/1.
We are grateful to V. Rittenberg and T. Klapwijk 
for useful discussions and to H. Weber for pointing out the 
experimental possibility of driving a ferromagnetic transition 
by a non-equilibrium transport voltage.

\end{document}